\documentclass{ws-procs975x65}

\begin{document}

\markboth{Kov\'acs, Gergely}{Hamiltonian theory of brane-world gravity}

\title{HAMILTONIAN THEORY OF BRANE-WORLD GRAVITY
\footnote{
Research supported by OTKA grants no. T046939 and TS044665, the J\'{a}nos
Bolyai Fellowships of the Hungarian Academy of Sciences, the Pierre Auger
grant 05 CU 5PD1/2 via DESY/BMF and the EU Erasmus Collaboration between the
University of Szeged and the University of Bonn. Z.K. and L.\'A.G. 
thank the organizers of the 11th Marcel Grossmann Meeting for support.} 
}

\author{ZOLT\'AN KOV{\'{A}}CS$^{\dagger \ddagger}$ and 
L\'ASZL\'O \'A. GERGELY$^{\ddagger}$}

\address{$\dagger$ Max-Planck-Institut f{\"u}r Radioastronomie,\\
Auf dem H\"ugel 69, D-53121 Bonn, Germany\\
$\ddagger$ Departments of Theoretical and Experimental Physics, University of Szeged,\\
D\'om t\'er 9, H-6720 Szeged, Hungary\\
\email{zkovacs@mpifr-bonn.mpg.de, gergely@physx.u-szeged.hu}}

\begin{abstract}
A brane-world universe consists of a 4-dimensional brane embedded into a
5-dimensional space-time (bulk). We apply the Arnowitt-Deser-Misner
decomposition to the brane-world, which results in a 3+1+1 break-up of the
bulk. We present the canonical theory of brane cosmology based on this
decomposition. The Hamiltonian equations allow for the study of any physical
phenomena in brane gravity. This method gives new prospects for studying the
initial value problem, stability analysis, brane black holes, cosmological
perturbation theory and canonical quantization in brane-worlds.
\end{abstract}


\bodymatter

\section*{}

The Hamiltonian theory of the brane-world scenario is based on the foliation
of the 4-dimensional (4d) world-sheet (the brane) which is embedded into the
5-dimensional (5d) space-time manifold (the bulk $\mathcal{B}$). Since the
3-dimensional (3d) space-like slices of the foliation admit tangent bundles
of co-dimension 2 with respect to $\mathcal{B}$, the slices form a
two-parameter family of 3-spaces $\Sigma _{t\chi }$ with $t,\chi \in I\!\!R$
embedded in $\mathcal{B}$. While the parameter $t$ represents the
many-fingered time in the canonical formalism, a new parameter $\chi $
defines the position of the brane in the bulk. A common choice is at $\chi =0
$. 

The 3+1+1 decomposition of the 5d brane-world geometry allows one to express the 5d field equations in terms of 3d quantities. These gravitational variables in these picture are the three metric $g_{ab}$ describing the intrinsic geometry of the slices $\Sigma _{t~\chi =0}=\Sigma _{t}$, a vector field $M^{a}$ and a scalar field $M$. The vector and scalar quantities describe the contribution of the bulk-gravity. The extrinsic curvature of the leaves embedded in $\mathcal{B}$ is given by the second fundamental forms $K_{ab}$ and $L_{ab}$ , the normal fundamental forms $\mathcal{K}^{a}=\mathcal{L}^{a}$%
, and scalars $\mathcal{K}$ and $\mathcal{L}$ associated with the two
time-like and space-like normal vector fields $n^{a}$ and $l^{a}$ of $\Sigma
_{t}$. The quantities $K_{ab}$, $\mathcal{K}^{a}$ and $\mathcal{K}$ are
equivalent with the time-derivatives of $g_{ab}$, $M^{a}$ and $M$,
respectively, whereas $L_{ab}$ and $\mathcal{L}$ contain only pure spatial
derivatives of them \cite{GK}.

As a result of the decomposition, the 5d Einstein-Hilbert action 
\begin{equation}
S^{G}[{}^{(5)}g_{ab}]=\int d^{5}xL^{G}=\int d^{5}x\sqrt{-{}^{(5)}g}{}^{(5)}R
\label{S}
\end{equation}%
(${}^{(5)}g_{ab}$ is the bulk metric and $^{(5)}R$ its scalar curvature) can
be expressed in terms of the set $(g_{ab},~M^{a},~M;~K_{ab},~\mathcal{K}%
^{a},~\mathcal{K},~L_{ab},~\mathcal{L})$, the lapse function $N$ and the
non-vanishing components of the shift vector $N^{a}$. The component of the
shift vector associated with the extra dimension is set to zero by the
condition of the confinement of the matter fields on the brane. With the set 
$(g_{ab},M^{a},M)$ chosen as canonical coordinates of the vacuum bulk
gravity, we can express the Lagrangian in the action (\ref{S}) solely in the
terms of canonical coordinates and their time derivatives. Then we introduce
the momenta $(\pi ^{ab},p_{a},p)$ conjugated to the canonical coordinates,
such that the phase space of the 5d vacuum gravity is the set 
\begin{equation}
(g_{A};\pi ^{A}|~A=1,2,3):=(g_{ab},~M^{a},~M;~\pi ^{ab},~p_{a},~p)
\label{g_A}
\end{equation}%
with the abstract index $A$ defined as $g_{1}=g_{ab}$, $g_{2}=M^{a}$, etc.
In order to specify the possible states in the phase space, we first need
the evolution equations, then the vacuum constraint equations which restrict
these solutions. (In fact they restrict only the initial data. Once they are
imposed, the dynamics preserves the constraints.)

The Legendre transformation of the decomposed vacuum Lagrangian yields to
the Hamiltonian $H^{G}$ of the bulk gravity. This is a linear combination of
the super-Hamiltonian constraint $H_{\bot }^{G}$ and supermonentum
constraint $H_{a}^{G}$: 
\begin{equation}
H^{G}[g_{A},\pi ^{A};N,N^{a}]=NH_{\bot }^{G}[g_{A},\pi
^{A}]-N^{a}H_{a}^{G}[g_{A},\pi ^{A}]\;.  \label{H}
\end{equation}%
When inserting the Lagrangian 
\begin{equation}
L^{G}[g_{A},\pi ^{A};N,N^{a}]=\pi ^{A}\dot{g}_{A}-H^{G}[g_{A},\pi
^{A};N,N^{a}]\;,  \nonumber
\end{equation}%
into the action (\ref{S}) and extremizing it with respect to the canonical
variables (\ref{g_A}), the lapse function and the shift vector, we obtain
the equations of motion and the constraints of the 5d vacuum gravity.

The Poisson brackets of any pair of functions on the phase space, as
in the field theories, can be defined with the help of the functional
derivatives of the functions with respect to the canonical variables or
merely via the canonical commutation relations. Then the dynamical equations
of the bulk gravity lead to the forms 
\begin{eqnarray*}
\dot{g}_{A}(x,\chi ) &=&\{g_{A}(x,\chi ),H^{G}[\mathbf{N}]\}\;, \\
\dot{\pi}^{A}(x,\chi ) &=&\{\pi ^{A}(x,\chi ),H^{G}[\mathbf{N}]\}
\end{eqnarray*}%
for any $x\in \Sigma _{t\chi }$ and $\chi \in I\!\!R$, where the notation $%
H^{G}[\mathbf{N}]$ includes the smearing of the Hamiltonian density (\ref{H}%
) with $N$ and $N^{a}$.

When matter fields couple to gravity on the brane, we enlarge the phase
space of the 5d geometry with the canonical variables of the matter source.
After decomposing the stress-energy tensor in the matter action, the
Hamiltonian of the matter fields can be derived as well. Then the dynamical
and the constraint equations of gravity must be supplemented with the
contributions from matter, a procedure resulting in the time-evolution
equations and the constraints of the total system of gravity and matter.

The Hamiltonian formalism for brane-world scenarios gives a suitable
starting point for canonical quantization. In this approach the quantum
state of gravity should be described by a state functional $\Psi (t,x,\chi
;g_{A}]$ over the configuration space $(g_{A})$. This functional
incorporates not only the intrinsic 3-geometries of the leaves $\Sigma
_{t\chi }$, but also the brane-off contributions of the bulk gravity. Dirac
constraint quantization imposes either the vacuum constraints $H_{\bot }^{G}$
and $H_{a}^{G}$, or the constraints of the gravity coupled to matter on the
state functional as operator equations, restricting the possible states of
the system: 
\begin{eqnarray}
\widehat{H}_{\bot }^{G}(t,x,\chi ;g_{A},\pi ^{A}]\Psi (t,x,\chi ;g_{A}]
&=&0\;,  \nonumber \\
\widehat{H}_{a}^{G}(t,x,\chi ;g_{A},\pi ^{A}]\Psi (t,x,\chi ;g_{A}] &=&0\;.
\label{hatH}
\end{eqnarray}%
By inserting the explicit form of the super-Hamiltonian constraint into the
first equation, where the canonical momenta $\pi ^{A}$ are represented by
the operators $\widehat{\pi }^{A}=-i\delta /\delta g_{A}$, we obtain a
second order functional differential equation of the state functional with
respect to the canonical coordinates. The latter is the Wheeler-deWitt
equation of the brane-world gravity, which may be simplified by applying the
operator restriction of the super-momentum constraint from Eqs. (\ref{hatH}).

We expect that the quantization of simple models already done in the context
of general relativity, such as the mixmaster universe \cite{M} or the
Einstein-Rosen waves \cite{K}, is possible in the brane-world scenario
either. The latter general relativistic model illustrates, together with
other examples,\cite{BrK,HKG} that various procedures can be found
which transform the constraint equations into new constraints, such that the
momentum canonically conjugated to the time variable enters the new
super-Hamiltonian only linearly. Hence the quantization of these systems
leads to a functional Schr{\"{o}}dinger equation instead of the second order
functional differential equation. There is hope that similar procedures
based on the Hamiltonian formulation of brane-world gravity\cite{GK} will be
successful for various simple brane-world models as well.

\bibliographystyle{ws-procs975x65}
\bibliography{Kovacs-AT3}

\end{document}